# Model-Free Adaptive Control based on Modified Full-Form-Dynamic-Linearization

Feilong Zhang

[1] State Key Laboratory of Robotics, Shenyang Institute of Automation, Chinese Academy of Sciences, Shenyang 110016, China

*zhangfeilong@sia.cn

**Abstract:** Current model-free adaptive control (MFAC) method hasn't been analysed in linear system and is not straightforward for the practical engineers to understand accurately. This correspondence presents a family of MFAC based on a modified equivalent-dynamic-linearization model (EDLM), which facilitates to show the working principle of method more directly and objectively. Compared to the current work, i) the researches on MFAC focus on linear model, which is easy to understand its working behaviour; ii) the full-form EDLM is extended with unmeasured stochastic and measured disturbance, respectively. Then the controllers is modified correspondingly; iii) the stability analysis of system cannot be proved by current contraction mapping technique when the sign of leading coefficient of control input changes, therefore we prove it by analysing the function of the closed-loop poles. Several simulated examples are used to show the principles of this kind of method.

## 1. Introduction

Recently, there has been significant progress in MFAC for a family of discrete-time nonlinear system. The controller design is based on a kind of process model called EDLM whose coefficients compose the pseudo-gradient (PG) vector. The time-varying PG is online estimated by the input and output data of the system and can be classified by compact-form (CF) EDLM, partial-form (PF) EDLM and full-form (FF) EDLM. The MFAC controller is designed through solving the optimal solution of the combination of quadratic index function and the process model. The relationships among these family of process model is that the MFAC method designed by CF-EDLM or PF-EDLM can be incorporated into that by FF-EDLM. Therefore, this note only discuss the FFDL-MFAC in detail [1]-[17].

The conception of model-free method proposed for nonlinear system in [1]-[17] are not straightforward for majorities of operating engineers to understand accurately. To exhibit the principle of this kind of adaptive controller more clearly, we analyze this kind control method in linear systems. Since the fundamental tool of current MFAC is transforming the nonlinear system model to EDLM at each point according to the principle of Cauchy mean value theorem or Taylor series. It means that the EDLM, which represents a local linearization of a nonlinear process, and controller character with linear incremental form. Thus, the design of this kind of method essentially bases on the linear model. The adaptive nature of the current MFAC to the uncertainty and nonlinear system is achieved by combining this optimal linear controller with the online estimation on the basis of the certainty-equivalent principle. Therefore, the study on the MFAC should begin with the linear system for more easily mastering its essence.

On one hand, the key parameter λ choosing method of MFAC is just based on qualitative analysis. And [1]-[10] demonstrate that the λ should be big enough to guarantee the convergence of tracking error. To correct this conclusion and complete the quantitative analysis about the relationship between the dynamic characteristics of system and λ, we present a simple method for controller design and parameter choosing method. The notion is that we should determine the suitable control law and its adjustable parameters with aims to acquire the desired closed-loop poles, which also means that the closed-loop system should characterize the visibly desired behaviour. Furthermore, this simple manner shows a practical guideline for the application of this method and is useful for understanding this kind of adaptive control method as well.

On the other hand, the sign of leading coefficient of control input in PG estimated vector is presupposed unchanged and its estimated value will be reset into the initial value when the sign changes. On the contrary, we shows that the sign of estimated can be changed in accordance with the changes of the system in simulation. If the component of PG vector are reset according to [1], the actual meaning of estimate values might be changed. Therefore, we keep the estimate method working in its own way without resetting value. Moreover, the stability analysis method of the current MFAC is based on the sign of $\phi_{Ly+1}(k)$ and $\hat{\phi}_{Ly+1}(k)$ unchanged, which is crucial to the current contraction mapping technique. To that end, we prefer an easier and direct manner by analyzing the function of the closed-loop poles to address this problem.

Besides, the previous works on FF-EDLM-MFAC and PF-EDLM-MFAC are only researched on the disturbance-free cases, whilst there generally exist the unmeasured bounded noises or measurement disturbances in practical situation. To this end, we modify the FF-EDLM by introducing the unmeasured external stochastic disturbances and measurement disturbances, then a class of MFAC method is designed by a more general cost function with taking account of the influence of the unmeasured external stochastic disturbances and measurement disturbances. Additionally, we present the proof of this kind of modified model and analyze the stability of the system subjected to the noise and disturbance. It shows that the tracking precision of system controlled by modified method is theoretically proved better than that controlled by the current method.

The organization of the paper is as follows: In Section 2, the current EDLM and MFAC are reviewed, and the relationship between the LTI DARMA model and current dynamic linearization model is analyzed. Then the characteristic roots and stability of the system controlled by MFAC is analyzed and the simulations are presented. Section 3S and Section 4 have develops a range of possible modified MFAC in the same procedure with Section 2. Section 5 gives the conclusion.

## 2. EDLM and Design of MFAC

The current EDLM method as a basic knowledge for the MFAC controller design is reviewed in 2.1, and its fundamental assumptions and theorem are presented as follows. Then the relationship between the DARMA model and EDLM is discussed. The MFAC control design and its stability analysis are shown in 2.2.

### 2.1. EDLM

We consider the following discrete-time SISO system.
$$y(k+1) = f(y(k), \cdots, y(k-n_y), u(k), \cdots, u(k-n_u)) \quad (1)$$

where $f(\cdot) \in R$ represents the unknown function; $u(k)$ and $y(k)$ represents the input and output of the system at time $k$, respectively. And $n_u, n_y \in Z$ represent their orders.

Suppose that the nonlinear system (1) conforms to below assumptions:

*Assumption 1*: The partial derivatives of $f(\cdots)$ with respect to all variables are continuous.

*Assumption 2*: System (1) conforms to the following generalized Lipschitz condition.
$$|y(k_1+1) - y(k_2+1)| \leq b \|\boldsymbol{H}(k_1) - \boldsymbol{H}(k_2)\| \quad (2)$$

where, $\boldsymbol{H}(k) = \begin{bmatrix} \boldsymbol{Y}_{Ly}(k) \\ \boldsymbol{U}_{Lu}(k) \end{bmatrix} = [y(k), \cdots, y(k-L_y+1), u(k), \cdots, u(k-L_u+1)]^T$ is a

vector which consists of control input and output of system within the time window $[k-L_u+1, k]$ and $[k-L_y+1, k]$, respectively. Two integers $L_y (1 \leq L_y \leq n_y)$ and $L_u (1 \leq L_u \leq n_u)$ are named pseudo orders of the system. For more details about *Assumption 1* and *Assumption 2*, please refer to [1]. [2].

*Theorem 1*: Considering nonlinear system (1) satisfying *Assumptions 1* and *2*, if $\Delta \boldsymbol{H}(k) \neq 0$, $1 \leq L_y \leq n_y$, $1 \leq L_u \leq n_u$, then a time-varying vector $\boldsymbol{\phi}_L(k)$ named PG vector exists and system (1) can be transformed into:
$$\Delta y(k+1) = \boldsymbol{\phi}_L^T(k) \Delta \boldsymbol{H}(k) \quad (3)$$

with $\|\boldsymbol{\phi}_L(k)\| \leq b$ for any $k$, where

$\boldsymbol{\phi}_L(k) = \begin{bmatrix} \boldsymbol{\phi}_{Ly}(k) \\ \boldsymbol{\phi}_{Lu}(k) \end{bmatrix} = [\phi_1(k), \cdots, \phi_{Ly}(k), \phi_{Ly+1}(k), \cdots, \phi_{Ly+Lu}(k)]$,

$\Delta \boldsymbol{H}(k) = \begin{bmatrix} \Delta \boldsymbol{Y}_{Ly}(k) \\ \Delta \boldsymbol{U}_{Lu}(k) \end{bmatrix} = [\Delta y(k), \cdots, \Delta y(k-L_y+1), \Delta u(k), \cdots, \Delta u(k-L_u+1)]^T$.

And we define $\phi_{Ly}(z^{-1}) = \phi_1(k) + \cdots + \phi_{Ly}(k) z^{-Ly+1}$, $\phi_{Lu}(z^{-1}) = \phi_{Ly+1}(k) + \cdots + \phi_{Ly+Lu}(k) z^{-Lu+1}$, $z^{-1}$ is the backward-shift operator.

*Proof*: Refer to [1] for details.

For LTI DARMA model:
$$A(z^{-1}) y(k+1) = B(z^{-1}) u(k) \quad (4)$$

Where $A(z^{-1}) = 1 + a_1 z^{-1} + \cdots + a_m z^{-m}$, $B(z^{-1}) = b_1 + \cdots + b_n z^{-n+1}$, are polynomials in unit delay operator $z^{-1}$, and $n$, $m$ are the orders of the system. Letting (4)- $z^{-1}$ (4) yields
$$\Delta y(k+1) = \alpha(z^{-1}) \Delta y(k) + \beta(z^{-1}) \Delta u(k) \quad (5)$$

where,
$$\alpha(z^{-1}) = -a_1 - \cdots - a_m z^{-m+1}$$
$$\beta(z^{-1}) = b_1 + \cdots + b_n z^{-n+1}$$

Let $\phi_{Ly}(z^{-1}) = \alpha(z^{-1})$ and $\phi_{Lu}(z^{-1}) = \beta(z^{-1})$, (5) is rewritten as (3). This illustrates that the (3) can be expressed by (4) and indicates that the current EDLM can be incorporated into the DARMA model. More precisely, it also means that any EDLM can be expressed by DARMA, while not all the DARMA model can be expressed by the EDLM appropriately. Nevertheless, this cannot affect the superiority of the controller designed through this kind of incremental form of model.

### 2.2. Design of MFAC

We can rewrite (3) into (6).
$$y(k+1) = y(k) + \boldsymbol{\phi}_L^T \Delta \boldsymbol{H}(k) \quad (6)$$

The object is to design a controller that guarantees closed-loop stability and optimizes output tracking performance in the sense that:
$$J = |y^*(k+1) - y(k+1)|^2 = \min imum \quad (7)$$

Where, $\lambda$ is a positive weighted constant; $y^*(k+1)$ is the desired system output signal.

Substitute Equation (6) into Equation (7) and solve the optimization condition $\partial J / \partial \Delta u(k) = 0$, then we have:
$$\Delta u(k) = \frac{1}{\phi_{Ly+1}(k)} [y^*(k+1) - y(k) - \sum_{i=1}^{Ly} \phi_i(k) \Delta y(k-i+1) - \sum_{i=Ly+2}^{Ly+Lu} \phi_i(k) \Delta u(k+L_y - i+1)] \quad (8)$$

(8) is the optimal controller for (7). Then we change the coefficient $\frac{1}{\phi_{Ly+1}(k)}$ into $\frac{\phi_{Ly+1}(k)}{\lambda + \phi_{Ly+1}^2(k)}$ to prevent the denominator from being zero and not to influence the sign of this coefficient, and the controller will become (9).
$$\Delta u(k) = \frac{\phi_{Ly+1}(k)}{\lambda + \phi_{Ly+1}^2(k)} [y^*(k+1) - y(k) - \sum_{i=1}^{Ly} \phi_i(k) \Delta y(k-i+1) - \sum_{i=Ly+2}^{Ly+Lu} \phi_i(k) \Delta u(k+L_y - i+1)] \quad (9)$$

Controller (9) is also the solution of optimization $\partial J / \partial \Delta u(k) = 0$ of (10) in [1], [2] and it incorporates fluctuations of inputs, outputs and desired set points.
$$J = |y^*(k+1) - y(k+1)|^2 + \lambda |\Delta u(k)|^2 \quad (10)$$

Form (9) and (3), we can have
$$y(k) = \frac{\phi_{Ly+1}(k) \phi_{Lu}(z^{-1})}{T(z^{-1})} y^*(k) = \frac{z^{-1} \phi_{Ly+1}(k) \phi_{Lu}(z^{-1})}{T(z^{-1})} y^*(k+1)$$
(11)

$$u(k) = \frac{\phi_{L_y+1}(k)\left[1-z^{-1}\phi_{L_y}(z^{-1})\right]}{T(z^{-1})} y^*(k+1) \quad (12)$$

Where,
$$T(z^{-1}) = \lambda(1-z^{-1})\left[1-z^{-1}\phi_{L_y}(z^{-1})\right] + \phi_{L_y+1}(k)\phi_{L_u}(z^{-1}) \quad (13)$$

is the function of the closed-loop poles. And we can choose the locations of desired closed-loop poles by tuning the λ.

In [1], the tuning parameters $\rho_i \ (i=1,\cdots,L_y+L_u)$ are introduced to make the controller more flexible to settle the problem of modelling inaccuracy. However, this set of parameters may not always be suitable for the subsequent changes of actual system model or the strong nonlinearity of system.

For example: How can we design the MFAC to control the following nonlinear model.

$$y(k+1) = \begin{cases} 0.6y(k)-0.1y(k-1)+1.8u(k)-1.8u^2(k)+0.6u^3(k) \\ -0.15u(k-1)+0.15u^2(k-1)-0.05u^3(k-1) \quad 0<k<200 \\ y(k)+y(k-1)+u(k)+u(k-1) \quad\quad 200<k<400 \end{cases}$$

The model in the time of [0,200] is choose from [9], with the parameters chosen $L_y=2$, $L_u=2$ and $\rho_1=\rho_2=\rho_3=\rho_4=0.82$. However, the tuning parameters $\rho_i \ (i=1,\cdots,L_y+L_u)$ will change the meaning of the controller for the optimal cost index and affect the adaptive nature of the controller in the time of [200,400]. Therefore, we keep controller working on its own manner without introducing parameters $\rho_i$. In other words, the attention should, above all, be focused on how to choose the proper orders of model in each period and on the precision of parameters identification.

*Simulations:*

*Example* 1: In this example, the following discrete-time SISO linear structure-varying system is considered.

$$y(k+1) = \begin{cases} 0.2y(k)+0.8y(k-1)-0.5u(k) \\ +0.3u(k-1)+0.2u(k-2) \quad 0<k<350 \\ -0.4y(k)-0.5u(k)-0.2u(k-1) \quad 351<k<700 \end{cases} \quad (14)$$

The desired output trajectory is
$$y^*(k+1) = 5\times(-1)^{round(k/80)}, 1\leq k \leq 400$$

The controller parameters and initial setting for MFAC (6) and Astrom's minimum variance controller MVC are listed in Table I, and all of them should be identical. The estimate algorithm is adopted with the projection algorithm in [1] with tuning parameters η and μ.

**Table 1** Parameter Settings for MFAC and MVC

| Parameter | MFAC (6) | MVC |
|---|---|---|
| Order | $L_y=1$, $L_u=2$ | $n_a=1$, $n_b=2$ |
| η, μ | 3; 1 | 3; 1 |
| Initial value $\hat{\phi}_L(1)$ | [-0.8, -0.5, -0.2] | [-0.8, -0.5, -0.2] |
| u(0:6) | (0,0,0,0,0,0) | (0,0,0,0,0,0) |
| y(0:5) | (0,0,0,0.5,0.2) | (0,0,0,0.5,0.2) |

From Fig. 1, we can see that the tracking performance of system controlled by MFAC in (8) is very well, we almost cannot see the delay and overshoot at the time of [0, 350]. While there exists the static error in the system controlled by MVC. Since there exists integral action in the MFAC, the static error for square wave response is removed.

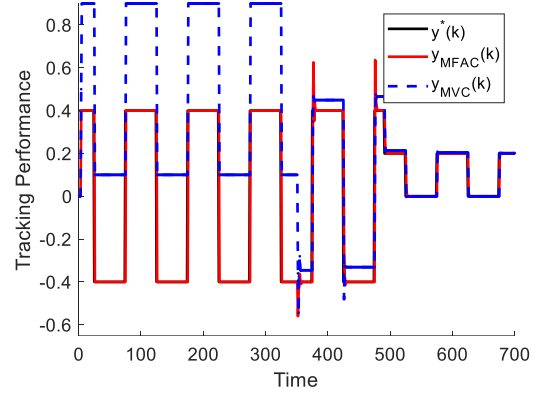

**Fig. 1** *Tracking performance*

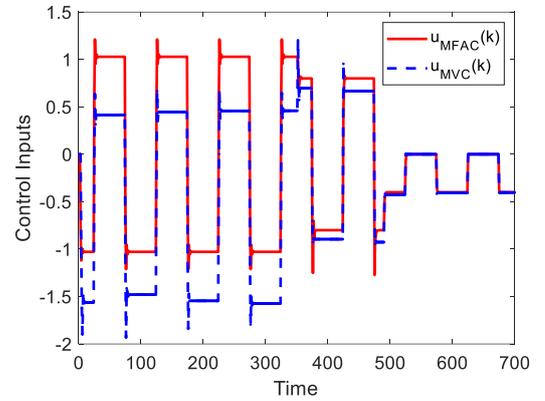

**Fig. 2** *Control input*

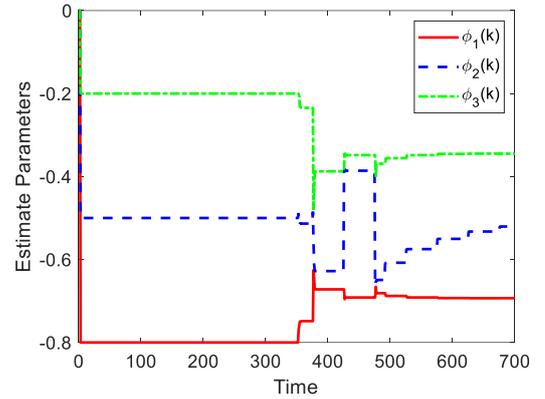

**Fig. 3** *Estimated value of PG*

*Example* 2: In this example, the following discrete-time SISO linear structure-varying system is given as.

$$y(k+1) = \begin{cases} -0.4y(k)-0.5u(k)-0.6u(k-1)+d_1 \quad 0<k<350 \\ 0.4y(k)+0.5u(k)+0.6u(k-1)+d_2 \quad 351<k<700 \end{cases} \quad (15)$$

where, $d$ is the disturbance. All the settings are in common with example 1, except $\hat{\phi}_L(1) = [-0.1,-0.1,-0.1]$ and λ= 0.2.

Case 1, $d_1=1$ and $d_2=100$.

Fig. 4 shows the tricking performance of the system controlled by MFAC in (9). Fig. 5 shows the control input. Fig. 6 shows the components of the PG estimation.

Case 2, $d_1 = 0$ and $d_2 = 0$.

Fig. 7 shows the tricking performance of the system controlled by MFAC in (9). Fig. 8 shows the components of the PG estimation.

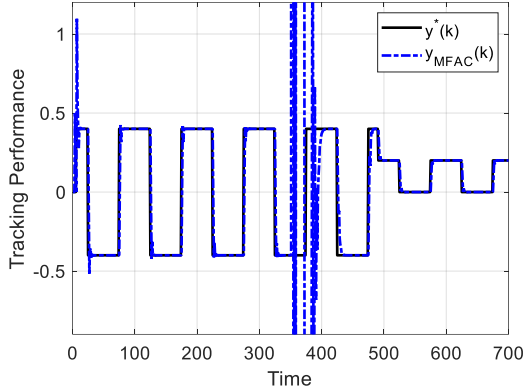

**Fig. 4** *Tracking performance*

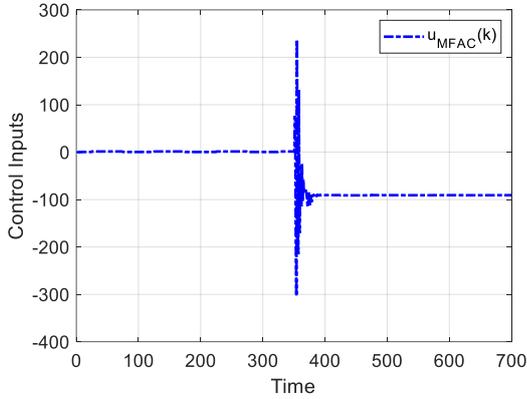

**Fig. 5** *Control input*

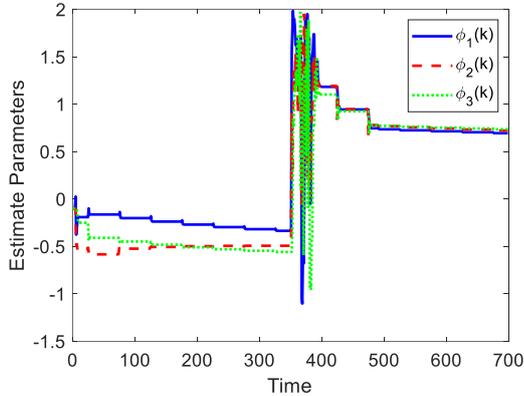

**Fig. 6** *Estimated value of PG*

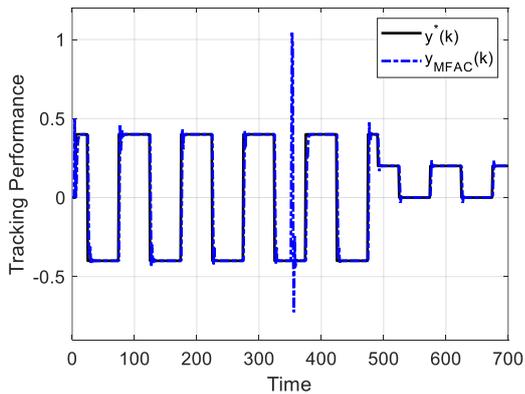

**Fig. 7** *Tracking performance*

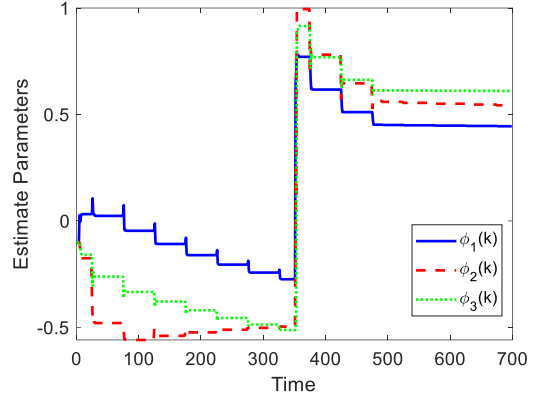

**Fig. 8** *Estimated value of PG*

From Fig. 4 and Fig. 7, we can see that the MFAC can remove the influence of constant disturbance to the static error of the system. Because this kind of incremental form controller is inherent based on the integer of tracking error. From Fig. 6 and Fig. 8, we can see that the sign of estimated $\hat{\phi}_{Ly+1}(k)$ is changed in the time of [350, 700]. If the components of PG vector are reset according to [1], the actual meaning of estimate values might be changed. In consequence, we keep the estimate method working in its own way without resetting value, aiming to validate the fact that all the signs of estimated components of PG are able to change. Nevertheless, the sign of $\hat{\phi}_{Ly+1}(k)$ unchanged is the crucial precondition of the current contraction mapping technique which is the current stability analysis method of the MFAC. To this end, we have analysed the stability of system by the means of the function of the closed-loop poles (13).

### 3. EDLM modified with Unmeasured Stochastic Disturbances and Design of General MFAC

In this section, 3.1 gives the EDLM with disturbance as a basic knowledge for the general MFAC design. Then the relationship between the ARMAX model and the modified EDLM is discussed. The general MFAC control law is designed with its stability analysis in 3.2.

#### 3.1. EDLM Modified with Unmeasured Stochastic Disturbance

We consider the following discrete-time SISO system:
$$y(k+1) = f(y(k), \cdots, y(k-n_y), u(k), \cdots, u(k-n_u) \\ , w(k), \cdots, w(k-n_w)) + w(k+1) \quad (16)$$

where $f(\cdot) \in R$ represents the unknown function, $n_y$, $n_u$, $n_w \in Z$ represent the unknown orders of the output $y(k)$, input $u(k)$ and the uncorrelated random sequence of zero mean disturbance (or noise) $w(k)$ of the system at time $k$, respectively.

Suppose that the nonlinear system (16) conforms to below assumptions:

*Assumption 3*: The partial derivatives of $f(\cdots)$ with respect to all its variables are continuous.

*Assumption 4*: System (16) conforms to the following generalized Lipschitz condition.
$$|y(k_1+1) - y(k_2+1)| \le b \|\boldsymbol{H}(k_1) - \boldsymbol{H}(k_2)\| + |w(k_1+1) - w(k_2+1))| \quad (17)$$

Where,
$$H(k) = \begin{bmatrix} Y_{Ly}(k) \\ U_{Lu}(k) \\ W_{Lw}(k) \end{bmatrix} = [y(k),\cdots,y(k-L_y+1),$$
$$u(k),\cdots,u(k-L_u+1), w(k),\cdots,w(k-L_w+1)]^T$$
is a vector which consists of control input, output and disturbance of system within the time window $[k-L_u+1,k]$, $[k-L_y+1,k]$ and $[k-L_w+1,k]$, respectively. Three integers $L_y(1\le L_y\le n_y)$, $L_u(1\le L_u\le n_u)$ and $L_w(0\le L_w\le n_w)$ are the pseudo orders of the system.

*Assumption 5*: $w(k)$ is an uncorrelated random sequence of zero mean disturbing the system with $E[\Delta w(k)^2]=\sigma^2$.

*Theorem 2*: Given system (16) such that *Assumptions 3, 4* and *5*, there must exist a time-varying vector $\phi_L(k)$ named PG vector; when $\Delta H(k)\ne 0$, $1\le L_y\le n_y$, $1\le L_u\le n_u$ and $1\le L_w\le n_w$, system (16) can be rewritten into the FF- EDLM with disturbance shown as follows
$$\Delta y(k+1) = \phi_L^T(k)\Delta H(k) + \Delta w(k+1) \quad (18)$$
with $\|\phi_L(k)\|\le b$ for any time $k$, where
$$\phi_L(k) = \begin{bmatrix} \phi_{Ly}(k) \\ \phi_{Lu}(k) \\ \phi_{Lw}(k) \end{bmatrix} = [\phi_1(k),\cdots,\phi_{Ly}(k),\phi_{Ly+1}(k),\cdots,\phi_{Ly+Lu}(k),$$
$$,\phi_{Ly+Lu+1}(k),\cdots,\phi_{Ly+Lu+Lw}(k)]^T$$
$$\Delta H(k) = \begin{bmatrix} \Delta Y_{Ly}(k) \\ \Delta U_{Lu}(k) \\ \Delta W_{Lu}(k) \end{bmatrix} = [\Delta y(k),\cdots,\Delta y(k-L_y+1),$$
$$\Delta u(k),\cdots,\Delta u(k-L_u+1),\Delta w(k),\cdots,\Delta w(k-L_w+1)]^T$$
And we define
$$\phi_{Ly}(z^{-1}) = \phi_1(k)+\cdots+\phi_{Ly}(k)z^{-Ly+1},$$
$$\phi_{Lu}(z^{-1}) = \phi_{Ly+1}(k)+\cdots+\phi_{Ly+Lu}(k)z^{-Lu+1}$$
$$\phi_{Lw}(z^{-1}) = \phi_{Ly+Lu+1}(k)+\cdots+\phi_{Ly+Lu+Lw}(k)z^{-Lw+1}$$

*Proof*: Please refer to Appendix.

*Assumption 6*: Suppose all the roots of the polynomial $1+z^{-1}\phi_{Lw}(z^{-1})=0$ within the open unit disk.

*Remark 2*: For ARMAX model:
$$A(z^{-1})y(k+1) = B(z^{-1})u(k) + C(z^{-1})\zeta(k) \quad (19)$$
Where, $\zeta(k)$ is uncorrelated random sequence of zero mean disturbance with variance $\dfrac{\sigma^2}{2}$.

$A(z^{-1}) = 1 + a_1 z^{-1} + \cdots + a_{na} z^{-na}$, $B(z^{-1}) = b_1 + \cdots + b_{nb} z^{-nb+1}$ and $C(z^{-1}) = 1 + c_1 z^{-1} + \cdots + c_{nc} z^{-nc}$ are polynomials in unit delay operator $z^{-1}$, and $n_a$, $n_b$, and $n_c$ are the orders of the system model. Letting (19)- $z^{-1}$ (19), we have
$$\Delta y(k+1) = \alpha(z^{-1})\Delta y(k) + \beta(z^{-1})\Delta u(k) + \gamma(z^{-1})\Delta \zeta(k+1) \quad (20)$$
Where
$$\alpha(z^{-1}) = -a_1 - \cdots - a_{na}z^{-na+1}$$
$$\beta(z^{-1}) = b_1 + \cdots + b_{nb}z^{-nb+1}$$
$$\gamma(z^{-1}) = 1 + c_1 z^{-1} + \cdots + c_{nc}z^{-nc}$$

Then letting $\phi_{Ly}(z^{-1}) = \alpha(z^{-1})$, $\phi_{Lu}(z^{-1}) = \beta(z^{-1})$ and $(1+\phi_{Lw}(z^{-1}))\Delta w(k+1) = \gamma(z^{-1})\Delta\zeta(k+1)$, we can obtain (18). This illustrates that the (18) can be expressed by (19).

### 3.2. Design of General MFAC Considering Unmeasured Stochastic Disturbance

We can rewrite (18) into (21).
$$y(k+1) = y(k) + \phi_L^T(k)\Delta H(k) + \Delta w(k+1)$$
$$= y(k) + \phi_{Ly}^T(z^{-1})\Delta y(k) + \phi_{Lu}^T(z^{-1})\Delta u(k) \quad (21)$$
$$+ \phi_{Lw}^T(z^{-1})\Delta w(k) + \Delta w(k+1)$$

The objective is to design a controller that guarantees the stability of closed-loop system and optimizes tracking error in the sense that:
$$J = |y^*(k+1) - y(k+1)|^2 = minimum \quad (22)$$
Where, $y^*(k+1)$ is the desired system output signal.

Suppose polynomial $G(z^{-1}) = g_0 + g_1 z^{-1} + \cdots + g_{ng}z^{-ng}$ ($n_g = n_a - 1$) is determined to satisfy the equation.
$$z^{-1}G(z^{-1}) = z^{-1}\phi_{Lw}(z^{-1}) + z^{-1}\phi_{Ly}(z^{-1}) \quad (23)$$

From (23) and (18), we have
$$(1+z^{-1}\phi_{Lw}(z^{-1}))\Delta y(k+1) = \left(\phi_{Lw}(z^{-1})+\phi_{Ly}(z^{-1})\right)\Delta y(k)$$
$$+\phi_{Lu}(z^{-1})\Delta u(k) + (1+z^{-1}\phi_{Lw}(z^{-1}))\Delta w(k+1) \quad (24)$$

(24) may be rewritten as (25).
$$y(k+1) = y(k) + \frac{\left(\phi_{Lw}(z^{-1})+\phi_{Ly}(z^{-1})\right)\Delta y(k) + \phi_{Lu}(z^{-1})\Delta u(k)}{(1+z^{-1}\phi_{Lw}(z^{-1}))}$$
$$+\Delta w(k+1) \quad (25)$$

Since $\Delta w(k+1)$ is supposed unmeasured, we let
$$y^*(k+1) = y(k) + \frac{\left(\phi_{Lw}(z^{-1})+\phi_{Ly}(z^{-1})\right)\Delta y(k) + \phi_{Lu}(z^{-1})\Delta u(k)}{(1+z^{-1}\phi_{Lw}(z^{-1}))} \quad (26)$$

to have the minimum of (22). Then we have
$$J_{\min} = E\left[|y^*(k+1)-y(k+1)|^2\right]$$
$$= E\left[(\Delta w(k+1))^2\right] = \sigma^2 \quad (27)$$

We can see that (27) is smaller than the index of current MFAC with $J_{current\ MFAC} = \left[1+\sum_{i=1}^{Lw}\phi_i^2(k)\right]\sigma^2$.

We may rewrite (26) into (28).
$$\left(\phi_{Lw}(z^{-1})+\phi_{Ly}(z^{-1})\right)\Delta y(k) + \phi_{Lu}(z^{-1})\Delta u(k)$$
$$= \left(1+z^{-1}\phi_{Lw}(z^{-1})\right)\left[y_d(k+1)-y(k)\right] \quad (28)$$

Then (28) may be rewritten as (29).

$$\Delta u(k) = \frac{1}{\phi_{Ly+1}(z^{-1})} \Big[ \big(1+z^{-1}\phi_{Lw}(z^{-1})\big)\big[y_d(k+1)-y(k)\big]$$
$$-\big(\phi_{Lw}(z^{-1})+\phi_{Ly}(z^{-1})\big)\Delta y(k) - \big(\phi_{Lu}(z^{-1})-\phi_{Ly+1}(z^{-1})\big)\Delta u(k)\Big]$$
$$= \frac{1}{\phi_{Ly+1}(z^{-1})} \Bigg[ \sum_{i=1}^{Lw} \phi_{Ly+Lu+i}(k)\big[y_d(k+1-i)-y(k-i)\big]$$
$$+ \big[y_d(k+1)-y(k)\big] - \sum_{i=2}^{Lu} \phi_{Ly+i}(k)\Delta u(k-i+1)$$
$$- \sum_{i=1}^{Lw} \phi_{Ly+Lu+i}(k)\Delta y(k-i+1) - \sum_{i=1}^{Ly} \phi_i(k)\Delta y(k-i+1) \Bigg]$$
(29)

(28) or (29) is the general form of MFAC in (9). In the case of $\phi_{Lw}(z^{-1})=0$, (29) will degenerate into (9).

Furthly, we choose the index function as follows
$$J_{\min} = \big[R(z^{-1})y^*(k+1) - P(z^{-1})y(k+1)\big]^2 + \big[\Lambda(z^{-1})\Delta u(k)\big]^2 \tag{30}$$

where, $P(z^{-1})=1+p_1 z^{-1}$, $R(z^{-1})=r_0+r_1 z^{-1}$, $\Lambda(z^{-1})=\lambda_0+\cdots+\lambda_{n\lambda}z^{-n\lambda+1}$ are the costing polynomials for the system output, desired set point and input.

Herein, we introduce the following Diophantine equation.
$$z^{-1}G(z^{-1}) = \big(1+z^{-1}\phi_{Lw}(z^{-1})\big)P(z^{-1}) - \big(1-z^{-1}\phi_{Ly}(z^{-1})\big) \tag{31}$$

Where, $G(z^{-1})$ derives from the polynomial identities. From (30), (21) and the optimization condition $\partial J/\partial \Delta u(k)$, we have:
$$\frac{\partial J}{\partial \Delta u(k)} = 2\big[Ry^*(k+1)-Py(k+1)\big]\frac{\partial y(k+1)}{\partial \Delta u(k)} + 2\frac{\partial J}{\partial \Delta u(k)}\big[\Lambda\Delta u(k)\big] \tag{32}$$

From [18], [19], (32) is minimized by choosing $\Delta u(k)$ such that
$$Ry^*(k+1) - Py(k+1) + \frac{\lambda_0}{\phi_{Ly+1}(k)}\Lambda\Delta u(k) = 0 \tag{33}$$

From (33), (31) and (18), we have
$$\left[\frac{\lambda_0}{\phi_{Ly+1}(k)}\big[1+z^{-1}\phi_{Lw}(z^{-1})\big]\Lambda(z^{-1}) + \phi_{Lu}(z^{-1})P(z^{-1})\right]\Delta u(k)$$
$$= \big[1+z^{-1}\phi_{Lw}(z^{-1})\big]\big[R(z^{-1})y^*(k+1)-P(z^{-1})y(k)\big] - G(z^{-1})\Delta y(k) \tag{34}$$

When $P(z^{-1})=1$, $R(z^{-1})=1$, $\Lambda=0$, the controller will degenerate into MFAC in (8). When $P(z^{-1})=1$, $R(z^{-1})=1$, $\phi_{Lw}(z^{-1})=0$, $\Lambda=\lambda_0=\sqrt{\lambda}$, it will degenerate into MFAC in (9).

From (34) and (21), we have
$$y(k) = \frac{\phi_{Lu}(z^{-1})R(z^{-1})}{T_3} y^*(k)$$
$$+ \frac{(1-z^{-1})\left(\big[1+z^{-1}\phi_{Lw}(z^{-1})\big]\frac{\lambda_0}{\phi_{Ly+1}(k)}\Lambda + \phi_{Lu}(z^{-1})\right)}{T_3} w(k) \tag{35}$$

$$u(k) = \frac{\big[1-z^{-1}\phi_{Ly}(z^{-1})\big]R(z^{-1})}{T_3} y^*(k+1)$$
$$- \frac{G+\big[1-z^{-1}\phi_{Ly}^T(z^{-1})\big]}{T_3} w(k) \tag{36}$$

where,
$$T_3 = \big[1-z^{-1}\phi_{Ly}(z^{-1})\big]\left[\frac{\lambda_0}{\phi_{Ly+1}(k)}\Lambda\right](1-z^{-1}) + P(z^{-1})\phi_{Lu}^T(z^{-1}) \tag{37}$$

is the function of the closed-loop poles.

The stability of the feedback system depends on the location of the roots of the polynomial equation shown as
$$\big[1-z^{-1}\phi_{Ly}(z^{-1})\big]\left[\frac{\lambda_0}{\phi_{Ly+1}(k)}\Lambda\right](1-z^{-1}) + P(z^{-1})\phi_{Lu}^T(z^{-1}) = 0 \tag{38}$$

Therefore, we can design the polynomials $P(z^{-1})$ and $\Lambda(z^{-1})$ to get the desired location of the roots of the polynomial equation to guarantee the stability of system and to acquire the desired system behaviours. According to [20]-[22], the BIBO stability of the system can also be guaranteed.

*Example* 3: In this example, the following discrete-time SISO linear system is considered.
$$y(k+1) = 1.7y(k) - 0.7y(k-1) + u(k) + 1.4u(k-1)$$
$$+ \xi(k) + 0.2\xi(k-1) \tag{39}$$

Where, $\xi(k)$ is uncorrelated zero-mean random sequence with variance 0.1. The desired output trajectory is
$$y^*(k+1) = 10 \times (-1)^{round(k/100)}, 1 \le k \le 400$$

The controller parameters and initial setting for general MFAC in (9) and MFAC in (34) are listed in Table 2, and all of them should be the same except $\Lambda(z^{-1})$. The estimation algorithm adopt the least square method in [20], [21] with parameter $\boldsymbol{P}(0) = 10^6 \boldsymbol{I}$.

**Table 2** Parameter Settings for MFAC in (9) and MFAC in (35)

| Parameter | General MFAC (34) | MFAC (9) |
|---|---|---|
| Order | $L_y=2$, $L_u=2$, $L_w=1$ | $L_y=2$, $L_u=2$ |
| $R(z^{-1})$, $P(z^{-1})$ | 1; 1 | 1; 1 |
| $\Lambda(z^{-1})$ | $0.5+0.2z^{-1}$ | 0.5 |
| Initial value $\hat{\phi}_L(1)$ | [0.001, 0.001, 0.001, 0.001] | |
| $u(0:3)$ | (0,0,0) | (0,0,0) |
| $y(0:2)$ | (0,0) | (0,0) |

Fig. 9 shows the tricking performance of the system controlled by both controllers. Fig. 10 shows the control input of both. Fig. 11 shows the components of the PG estimation. The performance indexes for both are shown in Table 3.

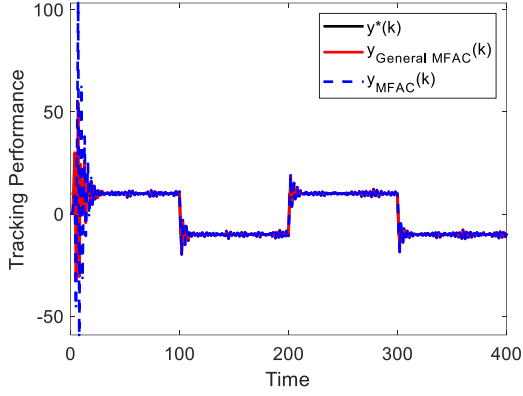

**Fig. 9** *Tracking performance*

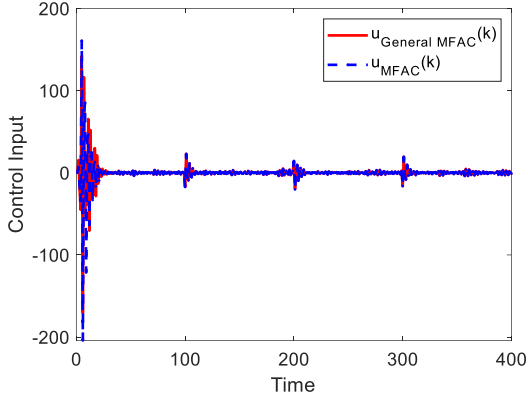

**Fig. 10** *Control input*

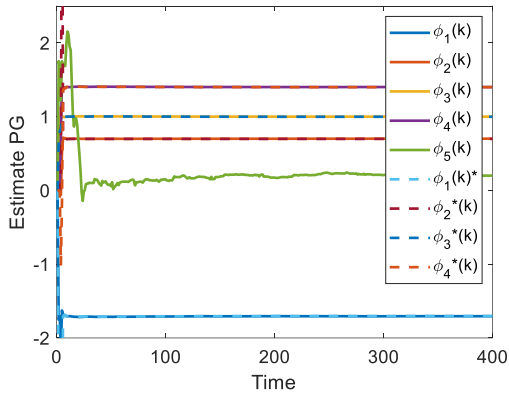

**Fig. 11** *Estimated value of PG*

**Table 3** Performance Indexes for MFAPC and MFAC

|  | Proposed MFAC | Current MFAPC |
|---|---|---|
| $eITAE = \sum_{k=1}^{N}|e(k)|^2$ | $7.71 \times 10^4$ | $7.91 \times 10^4$ |

From Fig. 9 and Table 3, we can see that the system controlled by general MFAC in (34) is slightly better than MFAC in (9) for less tracking error. Because the general MFAC has more tuning parameters to decide the location of the roots for the closed-loop poles equation and takes account of the influence of disturbance.

## 4. EDLM modified with measured disturbance and design of general MFAC

This section presents the EDLM modified with measured disturbance as a basic knowledge for the general MFAC controller design.

### 4.1. EDLM modified with Measurement Disturbance

We consider the following discrete-time SISO system:

$$y(k+1) = f(y(k),\cdots,y(k-n_y),u(k),\cdots,u(k-n_u)$$
$$v(k),\cdots,v(k-n_v),w(k),\cdots,w(k-n_w)) + w(k+1)$$
(40)

where $f(\cdot) \in R$ represents the unknown function, $n_y$, $n_u$, $n_v$, $n_w \in Z$ are supposed the unknown orders of the output $y(k)$, the input $u(k)$, the measured disturbance $v(k)$ and the uncorrelated random sequence of zero mean disturbance $w(k)$ at time $k$, respectively.

Suppose that the nonlinear system (40) conforms to the bellow assumptions:

*Assumption 7*: The partial derivatives of $f(\cdots)$ with respect to all variables are continuous.

*Assumption 8*: System (40) conforms to the following generalized Lipschitz condition.

$$|y(k_1+1) - y(k_2+1)| \leq b\|\boldsymbol{H}(k_1) - \boldsymbol{H}(k_2)\| + |w(k_1+1)) - w(k_2+1))|$$
(41)

Where,

$$\boldsymbol{H}(k) = \begin{bmatrix} \boldsymbol{Y}_{Ly}(k) \\ \boldsymbol{U}_{Lu}(k) \\ \boldsymbol{V}_{Lv}(k) \\ \boldsymbol{W}_{Lw}(k) \end{bmatrix} = [y(k),\cdots,y(k-L_y+1), u(k),\cdots,u(k-L_u+1)$$

$, v(k),\cdots,v(k-L_v+1), w(k),\cdots,w(k-L_w+1)]^T$

is a vector which consists of the system output, control input, measured disturbance and unmeasured disturbance within the moving time window $[k-L_y+1,k]$, $[k-L_u+1,k]$, $[k-L_v+1,k]$ and $[k-L_w+1,k]$, respectively. Four integers $L_y(1 \leq L_y \leq n_y)$, $L_u(1 \leq L_u \leq n_u)$, $L_v(0 \leq L_v \leq n_v)$ and $L_w(0 \leq L_w \leq n_w)$ represent pseudo orders of the system.

*Assumption 9*: $w(k)$ is an uncorrelated random sequence of zero mean disturbing the system with $E[\Delta w(k)^2] = \sigma^2$.

*Theorem 2*: Given system (1) satisfying *Assumptions 6*, *7* and *8*, there definitely has a time-varying PG vector $\boldsymbol{\phi}_L(k)$; If $\Delta \boldsymbol{H}(k) \neq 0$, $1 \leq L_y \leq n_y$, $1 \leq L_u \leq n_u$, $1 \leq L_v \leq n_v$ and $1 \leq L_w \leq n_w$, system (40) can be described as the following full-form-dynamic linearization data model with disturbance.

$$\Delta y(k+1) = \boldsymbol{\phi}_L^T(k)\Delta \boldsymbol{H}(k) + \Delta w(k+1)$$ (42)

with $\|\boldsymbol{\phi}_L(k)\| \leq b$ for any time $k$, where

$$\boldsymbol{\phi}_L(k) = \begin{bmatrix} \boldsymbol{\phi}_{Ly}(k) \\ \boldsymbol{\phi}_{Lu}(k) \\ \boldsymbol{\phi}_{Lv}(k) \\ \boldsymbol{\phi}_{Lw}(k) \end{bmatrix} = [\phi_1(k),\cdots,\phi_{Ly}(k),\phi_{Ly+1}(k),\cdots,\phi_{Ly+Lu}(k),$$

$\phi_{Ly+Lu+1}(k),\cdots,\phi_{Ly+Lu+Lv}(k),\phi_{Ly+Lu+Lv+1}(k),\cdots,\phi_{Ly+Lu+Lv+Lw}(k)]^T$

$$\Delta H(k) = \begin{bmatrix} \Delta Y_{Ly}^T(k) & \Delta U_{Lu}^T(k) & \Delta V_{Lv}^T(k) & \Delta W_{Lw}^T(k) \end{bmatrix}^T$$
$$= [\Delta y(k), \cdots, \Delta y(k-L_y+1), \Delta u(k), \cdots, \Delta u(k-L_u+1),$$
$$\Delta v(k), \cdots, \Delta v(k-L_v+1), \Delta w(k), \cdots, \Delta w(k-L_w+1)]^T$$

And we define $\phi_{Ly}(z^{-1}) = \phi_1(k) + \cdots + \phi_{Ly}(k)z^{-Ly+1}$,

$$\phi_{Lu}(z^{-1}) = \phi_{Ly+1}(k) + \cdots + \phi_{Ly+Lu}(k)z^{-Lu+1}$$
$$\phi_{Lv}(z^{-1}) = \phi_{Ly+Lu+1}(k) + \cdots + \phi_{Ly+Lu+Lv}(k)z^{-Lv+1}$$
$$\phi_{Lw}(z^{-1}) = \phi_{Ly+Lu+Lv+1}(k) + \cdots + \phi_{Ly+Lu+Lv+Lw}(k)z^{-Lw+1}.$$

*Proof*: Similar to Appendix.

*Assumption 10*: Suppose all the roots of the polynomial $1+z^{-1}\phi_{Lw}(z^{-1})=0$ within the open unit disk.

### 4.2. Design of General MFAC Considering Measured Disturbance

We can rewrite (42) into (43).
$$y(k+1) = y(k) + \phi_L^T(k)\Delta H(k) + \Delta w(k+1)$$
$$= y(k) + \phi_{Ly}^T(z^{-1})\Delta y(k) + \phi_{Lu}^T(z^{-1})\Delta u(k) \quad (43)$$
$$+ \phi_{Lv}^T(z^{-1})\Delta v(k) + \phi_{Lw}^T(z^{-1})\Delta w(k) + \Delta w(k+1)$$

The objective is to design a controller that guarantees closed-loop stability and optimizes output tracking performance in the sense that:
$$J_{\min} = \left[ R(z^{-1})y^*(k+1) - P(z^{-1})y(k+1) - S(z^{-1})v(k) \right]^2 \quad (44)$$
$$+ \left[ \Lambda(z^{-1})\Delta u(k) \right]^2$$

where, $P(z^{-1})=1+p_1z^{-1}$, $\Lambda(z^{-1})=\lambda_0+\cdots+\lambda_{n\lambda}z^{-n\lambda+1}$, $R(z^{-1})=r_0+r_1z^{-1}$, $S(z^{-1})=s_0+\cdots s_{ns}z^{-ns}$ are costing polynomials for the system output, input, desired set point and forward feedback of measured disturbance.

We introduce the following Diophantine equation.
$$z^{-1}G(z^{-1}) = \left(1+z^{-1}\phi_{Lw}(z^{-1})\right)P(z^{-1}) - \left(1-z^{-1}\phi_{Ly}(z^{-1})\right) \quad (45)$$

Where, $G(z^{-1})$ is derived from the polynomial identities. From (44) and the optimization condition $\partial J/\partial \Delta u(k)$, we have:
$$\frac{\partial J}{\partial \Delta u(k)} = 2\left[ Py(k+1) - Ry^*(k+1) + S(z^{-1})v(k) \right]\frac{\partial y(k+1)}{\partial \Delta u(k)}$$
$$+ 2\frac{\partial J}{\partial \Delta u(k)}\left[ \Lambda(z^{-1})\Delta u(k) \right]$$
(46)

From [18], [19], we know that (46) is minimized by choosing $\Delta u(k)$ such that
$$Ry^*(k+1) - Py(k+1) + S(z^{-1})v(k) + \frac{\lambda_0}{\phi_{Ly+1}(k)}\Lambda(z^{-1})\Delta u(k) = 0$$
(47)

From (47), (45) and (43), we have
$$\left[\frac{\lambda_0}{\phi_{Ly+1}(k)}\left[1+z^{-1}\phi_{Lw}(z^{-1})\right]\Lambda(z^{-1}) + \phi_{Lu}(z^{-1})\right]\Delta u(k)$$
$$= \left[1+z^{-1}\phi_{Lw}(z^{-1})\right]\left[R(z^{-1})y^*(k+1) - P(z^{-1})y(k)\right] - G(z^{-1})\Delta y(k)$$
$$- \left[\left[1+z^{-1}\phi_{Lw}(z^{-1})\right]S(z^{-1})v(k) + \phi_{Lv}(z^{-1})\Delta v(k)\right]$$
(48)

From (48) and (43), we have

$$y(k) = \frac{\phi_{Lu}^T(z^{-1})R(z^{-1})}{T_3} y^*(k)$$
$$+ \frac{\left( C(z^{-1})\frac{\lambda_0}{\phi_{Ly+1}(k)}\Lambda + \phi_{Lu}(z^{-1}) \right)}{T_3}\Delta w(k) \quad (49)$$
$$- \frac{\left[ \frac{\lambda_0}{\phi_{Ly+1}(k)}\Lambda\phi_{Lv}(z^{-1})\Delta - \phi_{Lu}(z^{-1})S(z^{-1}) \right]}{T_3}v(k-1)$$

$$u(k) = \frac{\left[ 1 - z^{-1}\phi_{Ly}^T(z^{-1}) \right]R(z^{-1})}{T_3} y^*(k+1)$$
$$- \frac{G + \left[ 1 - z^{-1}\phi_{Ly}^T(z^{-1}) \right]}{T_3} w(k)$$
$$- \frac{\left[ 1 - z^{-1}\phi_{Ly}^T(z^{-1}) \right]S(z^{-1}) + \phi_{Lv}^T(z^{-1})P(z^{-1})}{T_3} v(k)$$
(50)

Where,
$$T_3 = \left[1 - z^{-1}\phi_{Ly}^T(z^{-1})\right]\left[\frac{\lambda_0}{\phi_{Ly+1}(k)}\Lambda(z^{-1})\right](1-z^{-1}) + P(z^{-1})\phi_{Lu}^T(z^{-1})$$

is the function of the closed-loop poles. We can design the polynomials $P(z^{-1})$ and $\Lambda(z^{-1})$ to get the desired location of the roots of the polynomial equation. And we choose the $S(z^{-1})$ such that the following equation
$$\frac{\lambda_0}{\phi_{Ly+1}(k)}\Lambda(z^{-1})\phi_{Lv}(z^{-1})(1-z^{-1}) - \phi_{Lu}(z^{-1})S(z^{-1}) = 0 \quad (51)$$
to compensate the measured disturbance. According to [20]-[22], the BIBO stability of the system can easily be proved.

*Example* 4: In this example, the following discrete-time SISO linear system is considered.
$$y(k+1) = 1.7y(k) - 0.7y(k-1) + u(k) + 0.2u(k-1)$$
$$+ v(k) + 0.4v(k-1) + \xi(k) \quad (52)$$

where, $\xi(k)$ is uncorrelated zero-mean random sequence with variance 0.1; the measured disturbance is $v(k) = 5\sin(k/20)$.

The desired output trajectory is
$$y^*(k+1) = 10 \times (-1)^{round(k/100)}, 1 \leq k \leq 400$$

The controller parameters and initial setting for MFAC are $L_y=2$, $L_u=2$, $L_v=2$, $L_w=1$. The estimate method with parameters and initial settings are the same as Example 3. $\hat{\phi}_L(1) = [0.001, 0.001, 0.001, 0.001, 0.001, 0.001]$.

Fig. 12 shows the tricking performance of the system controlled by MFAC in (8). Fig. 13 shows the control input. Fig. 14 shows the components of the PG estimation.

Fig. 12 shows that the system, subject to the measured disturbance and controlled by MFAC, can track the desired trajectory. The fluctuation of control input in Fig. 13 indirectly reflects the measured disturbance. Therefore, the MFAC inherently has the advantageous ability in anti-disturbance to some degree.

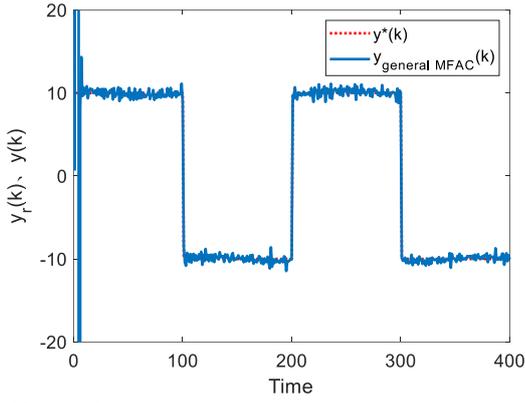

**Fig. 12** *Tracking performance*

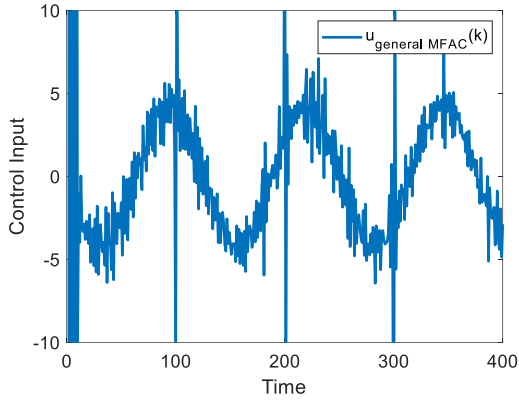

**Fig. 3** *Control input*

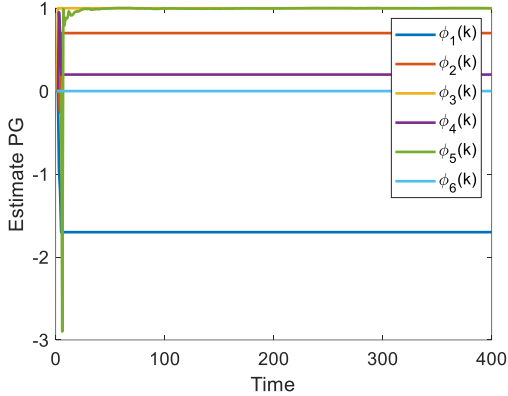

**Fig. 14** *Estimated value of PG*

## 5. Conclusion

In this note, we present a family of MFAC on their corresponding process model to show its working principle more directly and objectively. Additionally, the stability of system and parameter **λ** choosing method are analyzed by the function of closed-loop poles. Several simulated examples are used to validate the effectiveness and show the meaning of this family of method.

## 6. Appendix: Proof of Theorem 2

From (16), we have

$$\Delta y(k+1) =$$
$$f(y(k),\cdots,y(k-L_y+1),y(k-L_y)\cdots,y(k-n_y),u(k),$$
$$\cdots,u(k-L_u+1),u(k-L_u),\cdots,u(k-n_u),$$
$$w(k),\cdots,w(k-L_w+1),w(k-L_w),\cdots,w(k-n_w))+w(k+1)$$
$$-f(y(k-1),\cdots,y(k-L_y),y(k-L_y),\cdots,y(k-n_y),u(k-1),$$
$$\cdots,u(k-L_u),u(k-L_u),\cdots,u(k-n_u)$$
$$w(k-1),\cdots,w(k-L_w),w(k-L_w),\cdots,w(k-n_w))$$
$$+f(y(k-1),\cdots,y(k-L_y),y(k-L_y),\cdots,y(k-n_y),u(k-1),$$
$$\cdots,u(k-L_u),u(k-L_u),\cdots,u(k-n_u)$$
$$w(k-1),\cdots,w(k-L_w),w(k-L_w),\cdots,w(k-n_w))$$
$$-f(y(k-1),\cdots,y(k-L_y),y(k-L_y-1),\cdots,y(k-n_y-1),$$
$$u(k-1),\cdots,u(k-L_u),u(k-L_u-1),\cdots,u(k-n_u-1),$$
$$w(k-1),\cdots,w(k-L_w),w(k-L_w-1),\cdots,w(k-n_w-1))-w(k)$$
(53)

On the basis of Assumption 1 and Cauchy mean value theorem, Equation (53) becomes

$$\Delta y(k+1) = \frac{\partial f}{\partial y(k)}\Delta y(k)+\cdots+\frac{\partial f}{\partial y(k-L_y+1)}\Delta y(k-L_y+1)$$
$$+\frac{\partial f}{\partial u(k)}\Delta u(k)+\cdots+\frac{\partial f}{\partial u(k-L_u+1)}\Delta u(k-L_u+1)$$
$$+\frac{\partial f}{\partial w(k)}\Delta w(k)+\cdots+\frac{\partial f}{\partial w(k-L_w+1)}\Delta w(k-L_w+1)$$
$$+\psi(k)+\Delta w(k+1)$$
(54)

where,
$$\psi(k) \triangleq f(y(k-1),\cdots,y(k-L_y),y(k-L_y),\cdots,y(k-n_y),$$
$$u(k-1),\cdots,u(k-L_u),u(k-L_u),\cdots,u(k-n_u)$$
$$w(k-1),\cdots,w(k-L_w),w(k-L_w),\cdots,w(k-n_w))$$
$$-f(y(k-1),\cdots,y(k-L_y),y(k-L_y-1),\cdots,y(k-n_y-1),$$
$$u(k-1),\cdots,u(k-L_u),u(k-L_u-1),\cdots,u(k-n_u-1),$$
$$w(k-1),\cdots,w(k-L_w),w(k-L_w-1),\cdots,w(k-n_w-1))$$
(55)

$\frac{\partial f}{\partial y(k-i)}$, $0 \leq i \leq L_y-1$, $\frac{\partial f}{\partial u(k-j)}$, $0 \leq j \leq L_u-1$, and $\frac{\partial f}{\partial w(k-l)}$, $0 \leq l \leq L_w-1$ denote the partial derivative values of $f(\cdots)$ with respect to the $(i+1)$-th variable, the $(n_y+2+j)$-th variable and the $(n_y+n_u+3+l)$-th variable at some point within

$$[y(k),\cdots,y(k-L_y+1),y(k-L_y),\cdots,y(k-n_y),u(k),\cdots,$$
$$u(k-L_u+1),u(k-L_u),\cdots,u(k-n_u),$$
$$w(k),\cdots,w(k-L_w+1),w(k-L_w),\cdots,w(k-n_w)]$$

and $[(y(k-1),\cdots,y(k-L_y),y(k-L_y),\cdots,y(k-n_y),$
$u(k-1),\cdots,u(k-L_u),u(k-L_u),\cdots,u(k-n_u)$,
$w(k-1),\cdots,w(k-L_w),w(k-L_w),\cdots,w(k-n_w)]$
respectively.

We consider the following equation with the vector $\eta(k)$ for each time $k$:

$$\psi(k) = \boldsymbol{\eta}^T(k)\Delta \boldsymbol{H}(k) \tag{56}$$

Owing to $\|\Delta \boldsymbol{H}(k)\| \neq 0$, (56) must have at least one solution $\boldsymbol{\eta}^*(k)$. Let

$$\boldsymbol{\phi}(k) = \boldsymbol{\eta}^*(k) + [\frac{\partial f}{\partial y(k)}, \cdots, \frac{\partial f}{\partial y(k-L_y+1)},$$

$$\frac{\partial f}{\partial u(k)}, \cdots, \frac{\partial f}{\partial u(k-L_u+1)}, \frac{\partial f}{\partial w(k)}, \cdots, \frac{\partial f}{\partial w(k-L_w+1)}]^T \quad (57)$$

(54) can be described as follow:

$$\Delta y(k+1) = \boldsymbol{\phi}_L^T(k)\Delta \boldsymbol{H}(k) + \Delta w(k+1) \quad (58)$$

We finished the proof of *Theorem 2*.

## 7. Acknowledgments

This work was partially supported by the Science Fund for Creative Research Group of the National Natural Science Foundation of China (grant no. 91648204), the National Key Research and Development Program of China (grant no. 2017YFB130110).